\documentclass[a4paper,fleqn]{cas-dc}

\usepackage[numbers]{natbib}
\graphicspath{{images/}}

\def\tsc#1{\csdef{#1}{\textsc{\lowercase{#1}}\xspace}}
\tsc{WGM}
\tsc{QE}
\tsc{EP}
\tsc{PMS}
\tsc{BEC}
\tsc{DE}

\begin{document}
\let\WriteBookmarks\relax
\def\floatpagepagefraction{1}
\def\textpagefraction{.001}
\shorttitle{Versatile Cataract Fundus Image Restoration: Catintell}
\shortauthors{Z. Gong et~al.}

\title [mode = title]{Versatile Cataract Fundus Image Restoration Model Utilizing Unpaired Cataract and High-quality Images}                     

\tnotetext[1]{This work is supported by the Science and Technology Innovation Committee of Shenzhen-Platform and Carrier (International Science and Technology Information Center) \& Shenzhen Bay Lab under KCXFZ20211020163813019 and by the National Natural Science Foundation of China under 82000916. The authors declare no other conflict of interest. This study was performed in line with the principles of the Declaration of Helsinki. The Ethics Committee of Beijing Tongren Hospital, Capital Medical University approved and supervised the research process.}








\author[1]{Zheng Gong}[orcid=0000-0002-6220-3984]
\fnmark[1]
\ead{hudenjear@gmail.com}
\credit{Conceptualization of this study, Methodology, Software, Data curation, Writing - Original draft preparation}

\author[1]{Zhuo Deng}
\fnmark[1]
\ead{dengzhuo150@gmail.com}
\credit{Data curation, Writing - Original draft preparation}

\author[1]{Weihao Gao}
\ead{gwh20@mails.tsinghua.edu.cn}
\credit{Writing - Original draft preparation}

\author[2]{Wenda Zhou}
\ead{weiyizhouwenda@163.com}
\credit{Data curation}
\author[2]{Yuhang  Yang}
\ead{yyh63120@126.com}
\credit{Data curation}
\author[2]{Hanqing  Zhao}
\ead{zhq1214@126.com}
\credit{Data curation}

\author[1]{Zhiyuan Niu}
\ead{niuzy21@mails.tsinghua.edu.cn}
\credit{Data curation}

\author[2]{Lei Shao}
\ead{173625561@qq.com}
\credit{Data curation}

\author[2]{Wenbin Wei}
\ead{weiwenbintr@163.com}
\credit{Supervision}

\author[1]{Lan Ma}
\corref{cor1}
\credit{Supervision}

\cortext[cor1]{Corresponding author: 
  Email: malan@mails.tsinghua.edu.cn
  }

\address[1]{Tsinghua University Shenzhen International Graduate School, Guangzhou, Shenzhen 518055, China}
\address[2]{BeijingTongren Eye Center, Beijing Key Laboratory of Intraocular Tumor Diagnosis and Treatment, Beijing Ophthalmology \& Visual Sciences Key Lab, Beijing Tongren Hospital, Capital Medical University, Beijing, 100730, China}








\fntext[fn1]{These authors contribute equally.}


\begin{abstract}
Cataract is one of the most common blinding eye diseases and can be treated by surgery. However, because cataract patients may also suffer from other blinding eye diseases, ophthalmologists must diagnose them before surgery. The cloudy lens of cataract patients forms a hazy degeneration in the fundus images, making it challenging to observe the patient's fundus vessels, which brings difficulties to the diagnosis process. To address this issue, this paper establishes a new cataract image restoration method named Catintell. It contains a cataract image synthesizing model, Catintell-Syn, and a restoration model, Catintell-Res. Catintell-Syn uses GAN architecture with fully unsupervised data to generate paired cataract-like images with realistic style and texture rather than the conventional Gaussian degradation algorithm.
Meanwhile, Catintell-Res is an image restoration network that can improve the quality of real cataract fundus images using the knowledge learned from synthetic cataract images. Extensive experiments show that Catintell-Res outperforms other cataract image restoration methods in PSNR with 39.03 and SSIM with 0.9476. Furthermore, the universal restoration ability that Catintell-Res gained from unpaired cataract images can process cataract images from various datasets.
We hope the models can help ophthalmologists identify other blinding eye diseases of cataract patients and inspire more medical image restoration methods in the future.
\end{abstract}




\maketitle

\section{Introduction}
\label{sec:introduction}
The cataract is one of the most common causes of blindness. The World Health Organization estimates that cataracts will result in 40 million blindness in 2025 \cite{wang2016cataract}. Cataracts are typically caused by the deposition of proteins and form clouding of the lens in the eye. Cataracts usually develop with age but can also be caused by external factors such as trauma, diabetes, prolonged use of certain medications, or exposure to ultraviolet radiation. As cataracts grow, they can cause symptoms such as cloudy or blurred vision, faded colors, glare, poor night vision, and double vision. 

\begin{figure*}
  \centering
  \includegraphics[width=0.8\textwidth]{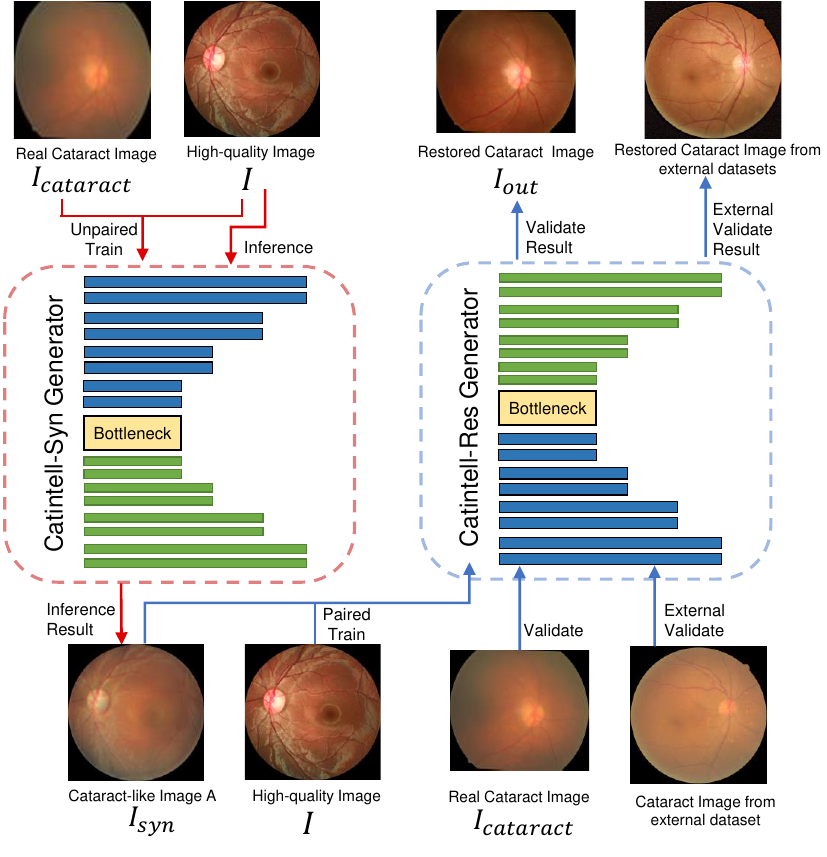}
  \caption{Catintell Model Workflow. We use two GAN models to generate synthetic cataract images and restore cataract images separately. The idea is to collect the information contained in real cataract images and let Catintell-Syn learn from it. Then Catintell-Res learns from synthetic data generated by Catintell-Syn and works on real cataract images from various datasets. Existing methods focus on learning from synthetic data generated by an old method\cite{41493}, which may not contain the features of real cataract images. But Catintell extracts features directly from real cataract images and applies them to real cataract image restoration.}
  \label{fig:CATflow}
\end{figure*}

Furthermore, cataracts also cause blurry clouding in retinal fundus photographing images and affect the diagnosis of other ophthalmic diseases through this method.
Fundus images have been expansively used in the fundus disease clinical diagnosis or computer-aided diagnosis systems. Since cataracts can cause lens opacity, the fundus images of cataract patients will suffer from fogging, blurring, and other degradation. It is challenging to make clinical diagnoses through low-quality cataract fundus images. Therefore, the low-quality fundus images could result in the risk of misdiagnosis and uncertainty in preoperative planning.  

Fundus image restoration can effectively solve the fundus image degradation caused by cataracts. Research in fundus image restoration has been carried out for many years. Traditional fundus image restoration methods~\cite{setiawan2013color,mitra2018enhancement,he2012guided,cheng2018structure} are mainly based on handcrafted priors. However, these methods achieve poor performance in clinical applications due to their limited prior knowledge or poor generalization ability. 

Recently, deep Convolutional Neural Networks(CNNs)\cite{chen2018gated,wang2018esrgan,li2022annotation,li2023generic} have been used in natural image restoration and achieved impressive results. CNNs have introduced into fundus image restoration due to the success in nature image restoration~\cite{zhao2019data,sengupta2020desupgan,shen2020modeling,raj2022novel,luo2020dehaze}.  
Meanwhile, the Transformer~\cite{9810184} has been introduced into fundus image restoration to address the limitations in capturing long-range dependencies and achieve remarkable performance. 
The advantage of the Transformer is capturing long-range dependencies. The effective combination of CNNs and Transformers may further improve the restoration performance of deep-learning models in cataract image restoration.

Since deep learning methods are mostly data-driven, existing cataract image restoration methods rely on a large number of cataracts and corresponding clear fundus image pairs. However, practical difficulties appear in cataract fundus image collecting. The degradation of cataract images is pathological, which means that clear images must be collected after surgeries to remove the clouding in the lenses. Nevertheless, collecting fundus images is not necessary after cataract surgery and may cause further damage to patients. Therefore, few cataract-clear image pairs were collected for now. Some cataract patients may have corresponding clear fundus images due to surgery follow-up, but, the long time gap of image collecting reduces the significance of these image pairs. There remains a lack of paired cataract images and clear images. 

To get training image pairs, the artificial degradation algorithm\cite{41493} was first brought out in 1989 and is used in many works even till now. Other models such as Gaussian filters~\cite{sengupta2020desupgan,shen2020modeling,li2022annotation} are designed to synthesize cataract-like images from high-quality (HQ) fundus images. However, these models can barely achieve good performance due to simple design. As shown in Fig.\ref{fig:exp-G}(b), these cataract-like images fundamentally differ from real clinical cataract images. 

In this paper, we set out to address the cataract image restoration problem. To alleviate the issue of lack of data, we propose a new cataract-like image synthesizing model, Catintell-Syn, which is a GAN model that uses fully unsupervised data to generate paired cataract-like images with realistic style and texture. Based on these simulated images, we develop a novel cataract fundus image restoration method, Catintell-Res, including a CNN-based generator and a Transformer-based discriminator. Specifically, the basic unit in the generator is the Dense Convolution Block(DCB),  which can capture local degradation features effectively. Unlike the generator, the basic unit of the discriminator is the Window-based Self-attention Block(WSB). The self-attention mechanism captures the non-local self-similarity and long-range dependencies, which can complement the shortcomings of CNNs. The Transformer-based discriminator can indirectly allow the generator to focus on non-local features through its classification ability with GAN architecture. Furthermore, the visual synthetic degradation comparison results show that the cataract-like images synthesized by our Catintell-Syn are closest to real cataract images in degradation style. Extensive experiments demonstrate that the Catintell-Res achieves remarkable performance in both synthetic cataract-like data and real cataract data. Finally, Catintell-Res is applied to real cataract images from various external datasets to verify its generalization performance and proved effective.

Our contributions can be summarized as follows:
\begin{enumerate}
    \item We propose a new image synthesizing method, Catintell-Syn, a deep learning model that only uses unpaired HQ and cataract images to generate realistic cataract images. 

    \item We develop a novel Transformer \& CNN-based method, Catintell-Res, for cataract fundus image Restoration. Considering the significant performance on multiple datasets. 

    \item Comprehensive quantitative and qualitative experiments demonstrate that our Catintell models outperform other state-of-the-art cataract image restoration algorithms.
\end{enumerate}

\section{Related Work}
\subsection{Fundus Image Restoration}

Traditional fundus image restoration and enhancement methods~\cite{setiawan2013color,mitra2018enhancement,he2012guided,cheng2018structure} are mainly based on hand-crafted priors. For example,  Setiawan $et~al.$ introduce CLAHE into fundus image enhancement\cite{setiawan2013color}.  Mitra $et~al.$ \cite{mitra2018enhancement} combines CLAHE with Fourier transform to enhance cataract images. He $et~al.$\cite{he2012guided} filter images as an edge-preserving smoothing operator and remove haze degradation efficiently. Cheng $et~al.$ \cite{cheng2018structure} propose a structure-preserving guided retinal image filtering (SGRIF) in fundus image restoration. However, these methods achieve poor performance in clinical applications due to their limited prior knowledge or poor generalization ability. 

CNN\cite{chen2018gated,wang2018esrgan,li2022annotation,li2023generic} have been used in natural image restoration and achieved impressive results. CNNs have introduced into fundus image restoration due to the success in nature image restoration~\cite{zhao2019data,sengupta2020desupgan,shen2020modeling,raj2022novel,luo2020dehaze}.  For instance, Zhao $\textit{et al}.$~\cite{zhao2019data} propose an end-to-end deep CNN to remove the lesions on the fundus images of cataract patients. Sourya $\textit{et al}.$~\cite{sengupta2020desupgan} , Shen $\textit{et al}.$~\cite{shen2020modeling}, and Raj $\textit{et al}.$~\cite{raj2022novel} customize different synthetic degradation models to simulate the degradation types in actual clinical practice better. Luo $\textit{et al}.$ report a two-stage dehazing algorithm,
which restores cataract fundus images under the supervision of segmentation~\cite{luo2020dehaze}. Li $\textit{et al}.$ ~\cite{li2022annotation} propose a network to annotation-freely restore
cataract fundus images (ArcNet). 

Meanwhile, the Transformer~\cite{9810184} has been introduced into fundus image restoration to address the limitations in capturing long-range dependencies and achieve remarkable performance. Deng $\textit{et al}.$ ~\cite{9810184} focus on real fundus image restoration and propose the first Transformer-based method (RFormer) in real fundus image restoration.

\subsection{Generative Adversarial Network}

Generative Adversarial Network (GAN) is firstly introduced in ~\cite{goodfellow2014generative} and has been proven successful in image synthesis~\cite{gong2019autogan,isola2017image,zhu2017unpaired}, and translation~\cite{isola2017image,zhu2017unpaired}. Subsequently, GAN is applied to image restoration and enhancement~\cite{wang2018esrgan,dbgan,sengupta2020desupgan,zhao2019data,jiang2021enlightengan}. For instance, Wang $\textit{et al}.$~\cite{wang2018esrgan} propose the ESRGAN in single image super-resolution. Zhang $\textit{et al}.$~\cite{dbgan} propose a new method that combines two GAN models, a learning-to-Blur GAN and learning-to-DeBlur GAN. Jiang $\textit{et al}.$~\cite{jiang2021enlightengan} focuses on low-light image enhancement and develop an unsupervised generative adversarial network(EnlightenGAN). Meanwhile, some works~\cite{jiang2021transgan,9810184} are dedicated to improving the underlying framework of GAN, such as replacing the traditional CNN framework with Transformer. Jiang $\textit{et al}.$~\cite{jiang2021transgan} propose the first Transformer-based GAN, TransGAN, for image generation. 

\begin{figure*}
  \centering
  \includegraphics[width=1\textwidth]{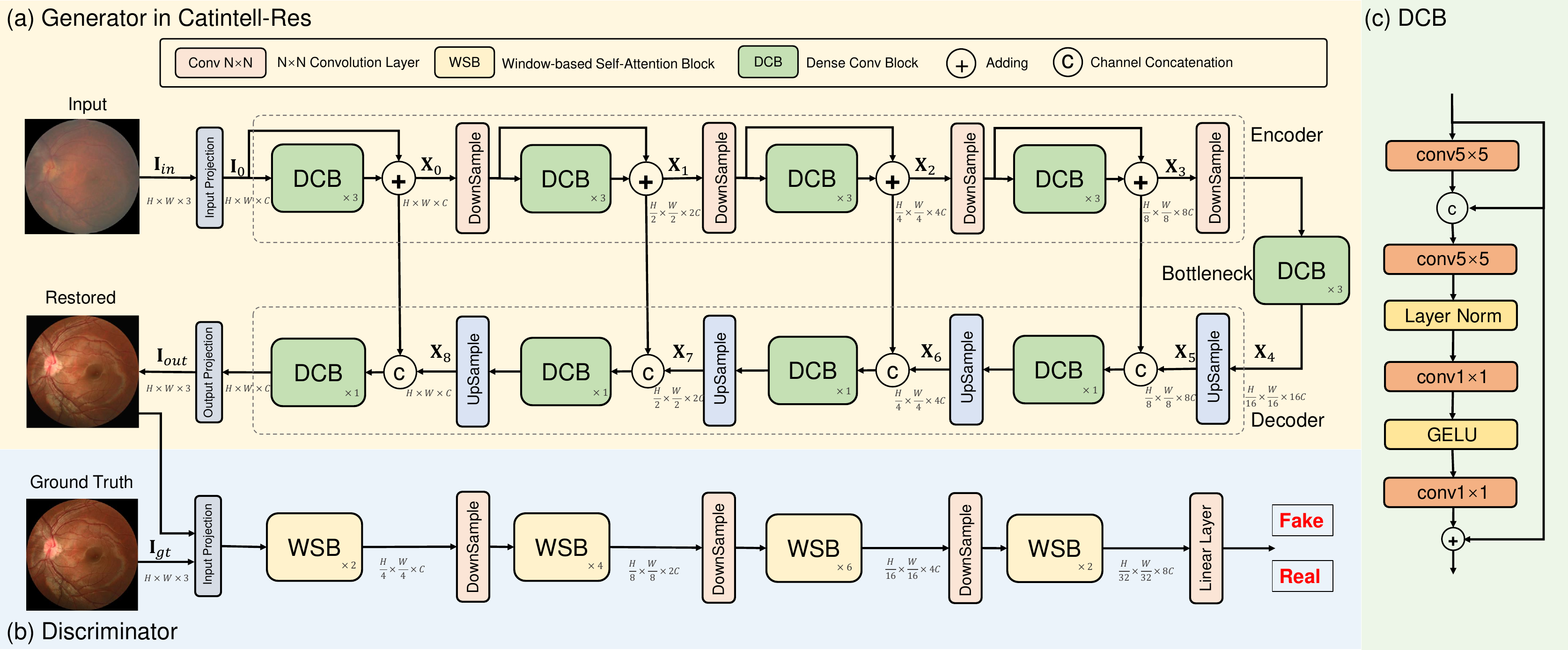}
  \caption{ The structure of the Catintell model. (a) The example model has a four-stage convolutional generator with downsampling and upsampling multiplier 2. (b) The discriminator of Catintell is a Transformer-based classifier and has four stages. (c) Detailed structure of the Dense Conv Block.}
  \label{fig:CAT-D}
\end{figure*}

\section{Methodology}
\subsection{Overview}

The Catintell model can be divided into two parts with similar structures: Catintell-Syn for image generation and Catintell-Res for cataract image restoration. Both of the Catintell models have the conditional GAN structure. 

Catintell-Syn receives HQ fundus images and generates synthetic cataract-like images of the same size. Catintell-Syn is trained with unaligned data from the Catintell dataset. Because cataract fundus images from the Catintell Image dataset have different sizes and height-width ratios, the HQ images are cropped to the same size and ratio to accelerate the convergence of Catintell-Syn. Meanwhile, this model receives low-quality cataract fundus images as input and outputs corresponding restored images. It can accept inputs of various sizes and height-width ratios and restore real cataract images. We use group convolution, internal small-range dense structures, and residual structures to improve performance.

After training with unpaired cataract data, we use Catintell-Syn to synthesize images highly similar to real cataract images. Then, these paired synthesized images are utilized to train Catintell-Res. This model follows a ``Pixel to Pixel" principle to restore fundus images with the same spatial size. Finally, the trained Catintell-Res can restore real cataract images from various sources.

\subsection{Catintell Model}

The structures of Catintell models are similar GAN architectures, therefore, here, we take the model used in the cataract image restoration stage, the Catintell-Res as an example, which is shown in Fig.\ref{fig:CAT-D}(a). Catintell-Res takes a cataract image $\mathbf{I}_{in} \in\mathbb{R}^{H\times W \times 3} $ as input. First, the input is processed by an input projection layer ($5\times5$ convolutional layer) to get the initial feature $\mathbf{I}_{0} \in\mathbb{R}^{H\times W \times C} $, where C is the feature dimension, and set to 32 in Catintell-Res. Then, the feature is encoded by three Dense Conv Blocks with a skip connection and downsampled with a convolutional layer. In the encoding stage, this operation is performed four times, and the spatial size of the feature can be denoted as $\mathbf{X}_{i} \in \mathbb{R}^{\frac{H}{ 2^{i+1}} \times \frac{W}{2^{i+1}} \times 2^{i+1}C}$. Here, \emph{i} = 0, 1, 2, 3 indicates the four stages. 
Afterward, the feature is processed by the bottleneck layers, another three Dense Conv Blocks, while its height, width, and channel are kept the same. Then, the feature is upsampled with four upsampling layers, each followed by one Dense Conv Block, and its spatial size is transferred to $\mathbf{X}_{i} \in \mathbb{R}^{\frac{H}{ 2^{8-i}} \times \frac{W}{2^{8-i}} \times 2^{8-i}C}$. Here, \emph{i} = 5, 6, 7, 8 indicates the four upsampling stages. There are also skip connections between encoding and decoding stages of the same spatial size. Finally, the feature is processed by an output projection layer ($5\times5$ convolutional layer) to provide the output image $\mathbf{I}_{out} \in\mathbb{R}^{H\times W \times 3} $.

The discriminator of Catintell-Res is a lightweight SWIN-Transformer~\cite{liu2021swin}. The structure of the discriminator is shown in Fig.\ref{fig:CAT-D}(b). We use BCE loss as GAN loss in Catintell-Res.

The structure of Catintell-Syn follows the same workflow, but its depth and width are lower. We shrink its size to reduce its encoding level and reduce its generation ability because cataracts only affect the lenses of the eyes and seldom cause vessel lesions in fundus images. If the generation ability of Catintell-Syn is too strong, we can observe some artifact lesions on the generated images. Therefore, the depth and width are optimized to 3 stages and 16 feature dimensions to degrade fundus images but not generate lesions.

\subsubsection{Conv Encoder}

In the encoding and decoding stages, the spatial size of feature maps does not change after processing by the Dense Conv Blocks or Conv Encoders. The structure of the Dense Conv Block is shown in Fig.\ref{fig:CAT-D}(d). It comprises two $5\times5$ convolutional layers and two $1\times1$ convolutional layers. There is layer normalization between $5\times5$ and $1\times1$ convolutional layers and GELU activation between $1\times1$ convolutional layers. The second $5\times5$ convolutional layer not only receives output from the layer ahead but also receives input with a skip connection to form a dense structure. A Conv Encoder contains three Dense Conv Blocks, whose structure is shown in Fig.\ref{fig:CAT-D}(c). There is a skip connection in its structure, which can accelerate its convergence and raise its performance.

\subsubsection{Catintell Loss Functions}
To formulate the loss functions, we denote the target HQ image A with $\mathbf{I}$, the input cataract-like image A with $\mathbf{I}_{syn}$, the real cataract image B with $\mathbf{I}_{cataract}$, the output restored image A with $\mathbf{I}_{out}$, the process of degradation generator with $Gen(\cdot)$, and the process of degradation discriminator with $Dis(\cdot)$.

\textbf{Pixel Loss}:
The pixel loss is a fundamental loss function in Catintell models, and we chose to apply it using the SmoothL1 loss function, $\mathcal{L}_{smoothL1}$, which is shown in the eq.\ref{eq:SL1}.

\begin{small}
\begin{equation}
\mathcal{L}_{smoothL1} =
\begin{cases}
 0.5\times(\mathbf{I}-\mathbf{I}_{out})^2,  -1<\mathbf{I}-\mathbf{I}_{out}<1\\
|\mathbf{I}-\mathbf{I}_{out}|-0.5,  otherwise \\
\end{cases}
\label{eq:SL1}
\end{equation}
\end{small}

\textbf{Fundus Perceptual Loss}:
Due to the massive difference between the fundus and common images, the perceptual loss shall be modified to suit fundus images. We retrained a VGG-19~\cite{simonyan2014very} network to formulate a perceptual loss specifically for fundus images, which is named Fundus Perceptual Loss (FPLoss). The VGG-19 is trained with the EyeQ~\cite{fu2019evaluation} dataset images with quality labels. The FPLoss works similarly to normal perceptual loss, and it can also give style loss. 

Using $\phi(\cdot)$ to denote the feature extractor of VGG-19 and $Gram(\cdot)$ to denote the Gram matrix calculation, if we assume the height and width of extracted feature maps are $H$ and $W$, the FPLoss, $\mathcal{L}_{fp}$, can be denoted as following eq.\ref{eq:CATFP}.

\begin{small}
\begin{equation}
\begin{split}
     \mathcal{L}_{fp} = \frac{1}{HW}& \sum_{i = 1}^{H} \sum_{j = 1}^{W}(\phi(\mathbf{I}_{out})(i,j) - \phi(\mathbf{I}_{cataract})(i,j))^2;\\
     \mathcal{L}_{fp-style} &= \frac{1}{HW} \sum_{i = 1}^{H} \sum_{j = 1}^{W}(\phi(Gram(\mathbf{I}_{out}))(i,j) - \\
     &\phi(Gram(\mathbf{I}_{cataract}))(i,j))^2
\end{split}
\label{eq:CATFP}
\end{equation}
\end{small}

\textbf{Identity Loss}:
The identity loss $\mathcal{L}_{ide}$ can ensure that the restoration model can keep fundus images unchanged when the input images are HQ images. (Contrary to the cataract image synthesis model Catintell-Syn, which can keep the style of input cataract images) The style and details of a real HQ image shall be kept the same after the process of this restoration model. With input $\mathbf{I}$, the processed image of the degradation branch is $Gen(\mathbf{I})$. The identity loss will calculate the pixel loss of $\mathbf{I}$ and $Gen(\mathbf{I})$. To be more specific, the pixel loss applied in the identity loss is SmoothL1 loss, therefore, the loss can be formulated as eq.\ref{eq:iden}.

\begin{small}
\begin{equation}
     \mathcal{L}_{Identity} =  \mathcal{L}_{smoothL1}(\mathbf{I},Gen(\mathbf{I}))
     \label{eq:iden}
\end{equation}
\end{small}

\textbf{GAN Loss}:
The discriminators in the Catintell models give predictions of possibility. Therefore, we use BCE loss as GAN loss of Catintell-Res. The calculating of $\mathcal{L}_{GAN}$ is shown in eq.\ref{eq:L-GAN}

\begin{small}
\begin{equation}
\begin{split}
    \mathcal{L}_{GAN} &= -(\mathbf{P}_Y log(\mathbf{P}_{out})+(1-\mathbf{P}_Y)log(1-\mathbf{P}_{out})), \\
    &where \mathbf{P}_Y=Dis(\mathbf{I}_{cataract}), \mathbf{P}_{out}= Dis(\mathbf{I}_{out})
    \label{eq:L-GAN}
\end{split}
\end{equation}
\end{small}

The overall losses of Catintell models can be formulated as follows eq.\ref{eq:L-G} and eq.\ref{eq:L-D}, and each loss is adjusted by loss weight before its loss symbol. The weight of each loss is adjusted according to its importance. The pixel loss is of low weight in the Catintell-Syn model, which has unpaired input images, but significant in the Catintell-Res model. Meanwhile, we use the high weight of the perceptual and style losses in Catintell-Syn but relatively low weight in Catintell-Res.

\begin{small}
\begin{equation}
    \mathcal{L}_{Syn} =  0.01\mathcal{L}_{smoothL1}+\mathcal{L}_{fp}+
    0.1\mathcal{L}_{ide}+0.1\mathcal{L}_{GAN}
    \label{eq:L-G}
\end{equation}
\end{small}

\begin{small}
\begin{equation}
    \mathcal{L}_{Res} =  \mathcal{L}_{smoothL1}+0.1\mathcal{L}_{fp}+
    0.01\mathcal{L}_{ide}+0.1\mathcal{L}_{GAN}
    \label{eq:L-D}
\end{equation}
\end{small}

\begin{figure}
  \centering
  \includegraphics[width=0.8\linewidth]{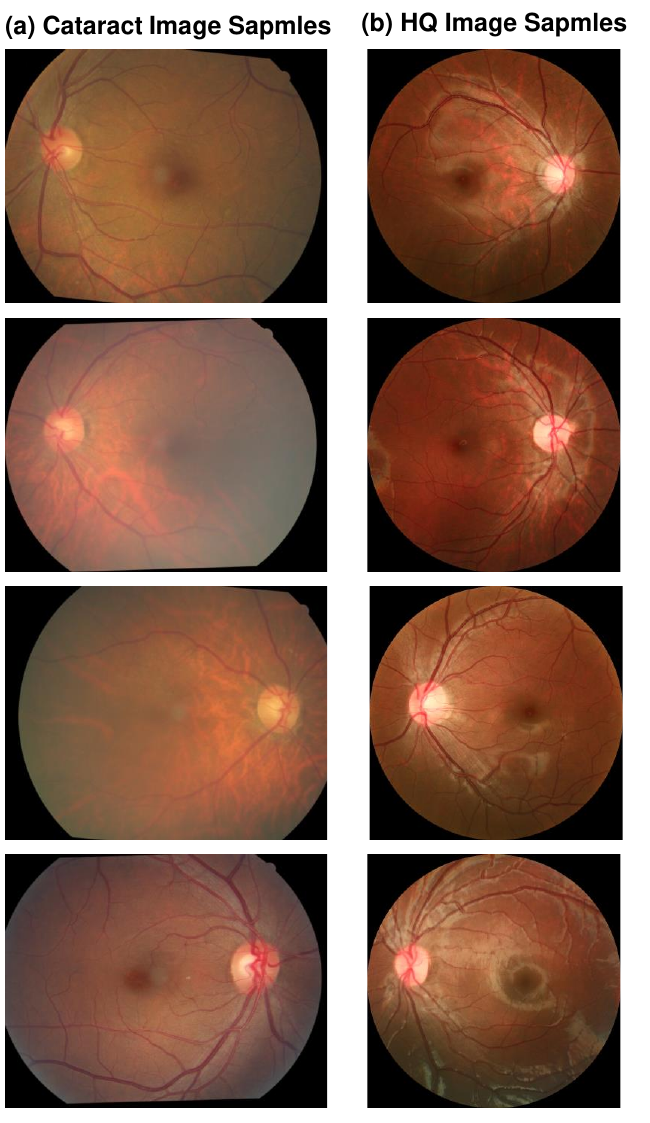}
  \caption{Sample of our Catintell Image dataset. (a) 2436 cataract images were collected in this dataset. (b) 1144 high-quality images were collected.}
  \label{fig:CAT-datasample}
\end{figure}

\section{Dataset and Experiments}

\subsection{Dataset}

To train and test Catintell models, we collected a dataset, named Catintell Image, containing 1144 HQ fundus images and 2436 cataract images from Beijing Tongren Hospital. Meanwhile, the 10-fold validation is also applied. Before training, collected images are randomly sampled 10\% as the validation set including 244 cataract images and 114 HQ images 10 times (we intend to create datasets without replication and absence, thus the last set contains 240 and 118 images), while the rest of these images are training set. Meanwhile, as mentioned above, the Catintell is a two-stage model, and the restoration of cataract images happens in the second stage which needs no clear or HQ images in the inference process. Therefore, we collected another 102 cataract images to examine the performance of Catintell in real cataract image restoration. There are some image samples shown in Fig.\ref{fig:CAT-datasample}. 

Besides the Catintell dataset, we also use two external datasets to validate the generality of the model. The ODIR\cite{odir} and an open-source Kaggle cataract dataset\cite{kaggle1} are experimented with to test the model's ability to enhance the quality of real cataract images.

The input fundus images are first resized to $768\times768$ and then randomly (paired) cropped to $256\times256$ patches. The spacial size of $768\times768$ can ensure details of original images are retained, and $256\times256$ is set for less GPU RAM usage and data augmentation. Meanwhile, all images are augmented using horizontal/vertical flipping. This data augmentation method is not applied to the validation and test stages to ensure consistent output and completeness of cataract images. The Catintell-Res model can enhance fundus images with different height-width ratios. Therefore, input image shapes in the validating and testing stages are flexible.

\subsection{Deployment Details}

During training, Catintell is applied with PyTorch version 1.10 and trained with CUDA version 11.7. We train each model for 80,000 iterations (equivalently 300 epochs) with the batch size 8 and learning rate $10^{-5}$ with cosine decay for all sub-models at first and apply a fine-tuning process with the same batch size and learning rate $10^{-6}$ with linear decay only for Catintell-Res models. The Adam~\cite{kingma2014adam} optimizer is applied with 1000 iterations warm up. All experiments are trained using a single NVIDIA Geforce RTX3090 GPU running for 10 hours to complete the training process.

The proposed Catintell model is an image restoration model, so we select the PSNR and SSIM as evaluation metrics. The optimization process of hyperparameters in Catintell models is demonstrated in the later part of the experiment section.

\begin{figure*}
  \centering
  \includegraphics[width=0.85\linewidth]{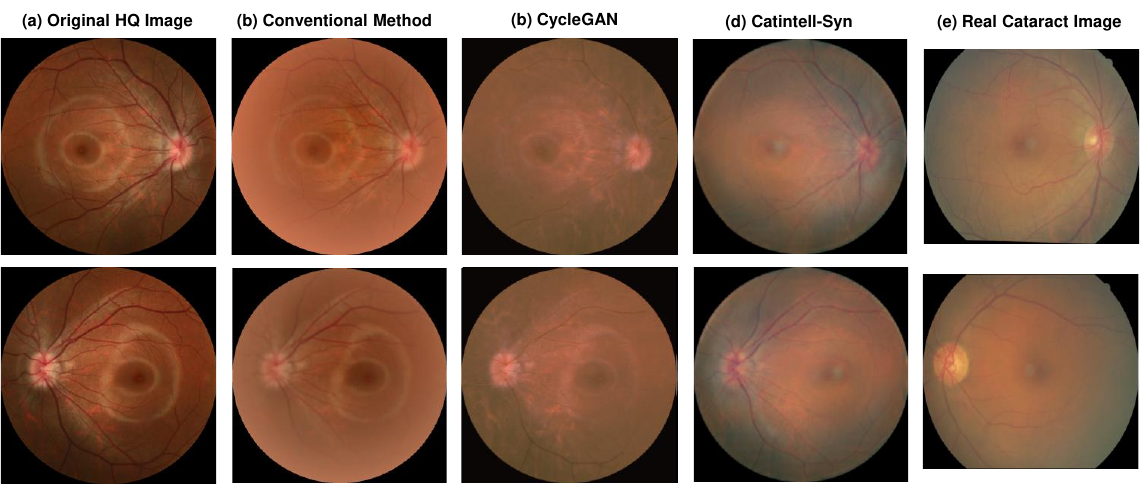}
  \caption{Result of degraded images from Catintell-Syn and traditional modeling method. (a) Source HQ fundus images. (b) Synthetic cataract fundus images using traditional method. (c) using CycleGAN. (d) using Catintell-Syn. (d) Real cataract fundus image samples. The images generated by Catintell-Syn are more similar to real cataract fundus images.}
  \label{fig:exp-G}
\end{figure*}

\subsection{Catintell-Syn Experiments}

We provide qualitative comparisons between Catintell-Syn, CycleGAN\cite{CycleGAN2017}, and the traditional degradation method\cite{41493}. The so-called 'traditional method' was first introduced in 1989\cite{41493}, and is utilized in many cataract restoration works mentioned above. Though this method can give promising output for various fundus images, it has trouble dealing with images with a height-width ratio other than 1:1. Moreover, this method follows a fixed algorithm workflow regardless of the difference between input fundus images and has outputs almost the same style. The CycleGAN is widely utilized in style transfer research, we also carry out experiments on this method. However, it did not achieve fair results on cataract images.

The results are shown in Fig.\ref{fig:exp-G}. It can be observed that the degradation style of Catintell-Syn is essentially consistent with real cataract images. Specifically, synthetic degradation closely matches real degradation in both location and severity. Severe degradation is observed in the blood vessels and macula area, while the optic disc region shows mild degradation.

\subsection{Catintell-Syn User Study}
To get real feedback from ophthalmologists, we conducted a user study to collect their opinions and rank cataract images synthesized by our Catintell-Syn model. In the study, we provide them with five images: real cataract images, HQ images, images from the conventional method, CycleGAN, and images from our Catintell-Syn model. There are ten sets of these image groups, and the images are given 10,8,6,4,2 scores corresponding to their ranks respectively. (This score setting means to get scores with a maximum of 100. Higher similarity to real cataract images results in higher scores.) The average results of three experienced ophthalmologists and three young ophthalmologists are summarized in Table \ref{tab:Gus}. 

\begin{table}
	\footnotesize
	\centering	
	\caption{User Study of Catintell-Syn}
	{
			\begin{tabular}{c c c c c c}
				\toprule
				 Image  & HQ & Conventional & CycleGAN & \textbf{Ours} & Real\\
				\midrule
				Rank Score &20&53&62&79&86\\
				\bottomrule
	\end{tabular}}
	\label{tab:Gus}
\end{table}

The score of images synthesized by Catintell-Syn is slightly lower than the real cataract images and obviously higher than images generated by the conventional method or CycleGAN. Therefore, we conclude that Catintell-Syn succeeds in synthesizing cataract images highly similar to real ones.

\begin{figure*}
  \centering
  \includegraphics[width=0.9\linewidth]{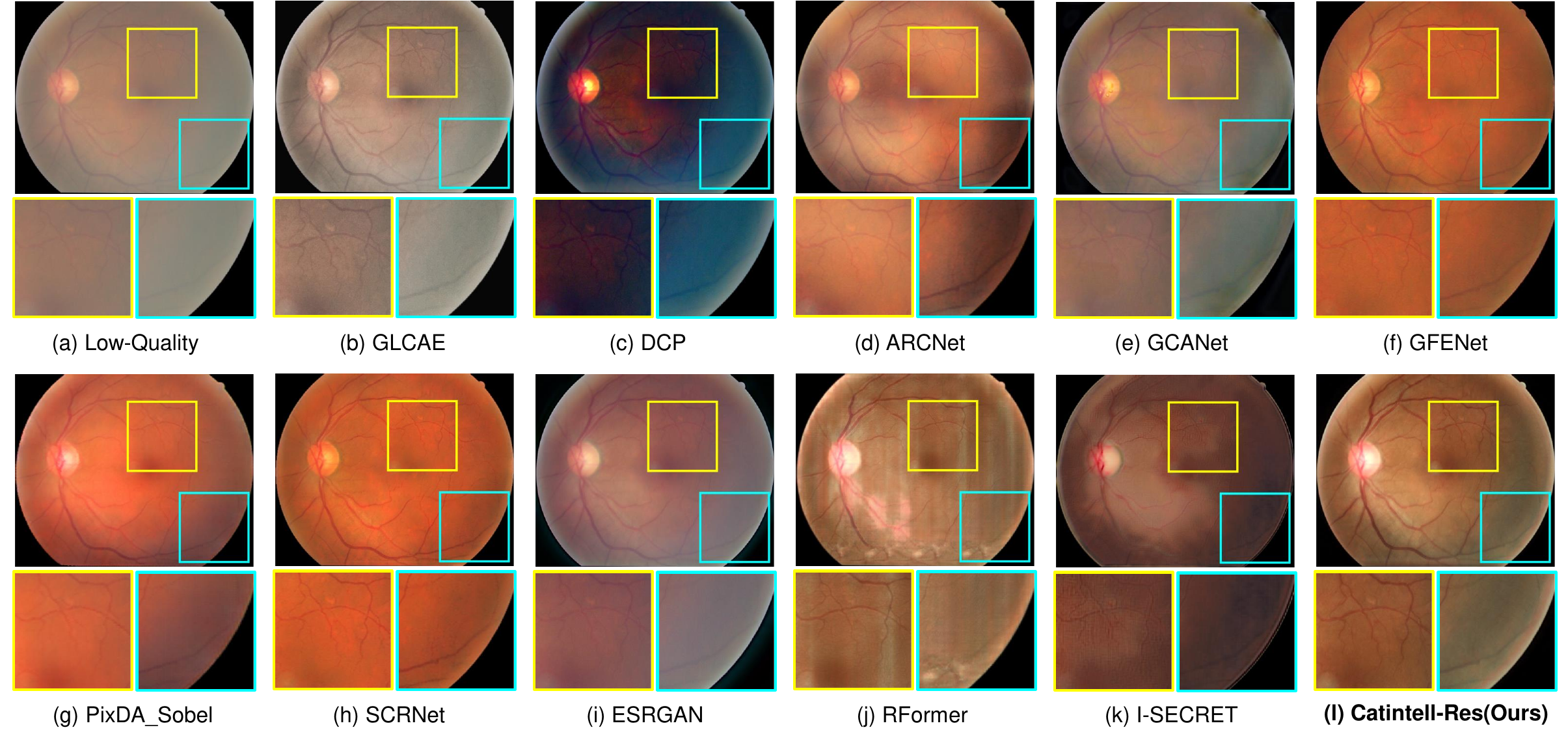}
  \caption{Restored real cataract image comparisons of Scene 1 on a test image of the Catintell Image dataset. Compared to other methods, the vessels around the macula in the restored image of Catintell-Res are finely enhanced. The overall style of this image is also maintained rather than changed to a dark/orange color.}
  \label{fig:real_Compare1}
\end{figure*}

\subsection{Catintell-Res Experiments}

The calculation of quantitive metrics requires paired images. However, as addressed in the introduction section, the difficulty of acquiring cataract-clear image pairs within a short interval hinders data collection. Therefore, to meter the performance of Catintell-Res models, we use the simulated cataract-HQ image pairs from the Catintell-Syn model. Moreover, the following models for comparison are also applied with the simulated cataract-HQ images to ensure fair comparisons. 

The GLCAE~\cite{tian2017global} and Dark channel prior(DCP)~\cite{8658661} are modeling methods and need no parameters. They usually follow the same work mode and apply the same modifications to different images. The ESRGAN~\cite{wang2018esrgan} and GCANET\cite{chen2018gated} are general image enhancement methods that are yet to be adapted to cataract image restoration. We retrained these models with cataract image pairs to get better results. The ARCNet\cite{li2022annotation}, pixDA Sobel\cite{li2021restoration}, SCRNET\cite{li2022annotation}, RFormer~\cite{9810184}, I-SECRET~\cite{cheng2021secret}, and GFENET\cite{li2023generic} are reported fundus image enhancement methods. The ARCNet, pixDA Sobel, and SCRNET use high-frequent information to enhance the restoration process, and RFormer uses Transformers to elevate its performance. These methods need algorithms to degrade the HQ images to get cataract-like images first and then restore the image. Therefore, they actually target a fixed fake cataract image-generating method but not the real style of cataract images. However, the Catintell can learn from the realistic cataract-like images which are proven better in the prior section. Meanwhile, the comparisons with general image restoration can also prove that Catintell is more suitable for the cataract image restoration task.

The results of quantitive metrics are shown in Table \ref{tab:CATsota}. We can observe from the results that Catintell-Res has a great ability for image restoration. It extravagantly outperforms other methods both in PSNR and SSIM through learning from the synthesized data.

\begin{figure*}
  \centering
  \includegraphics[width=0.9\linewidth]{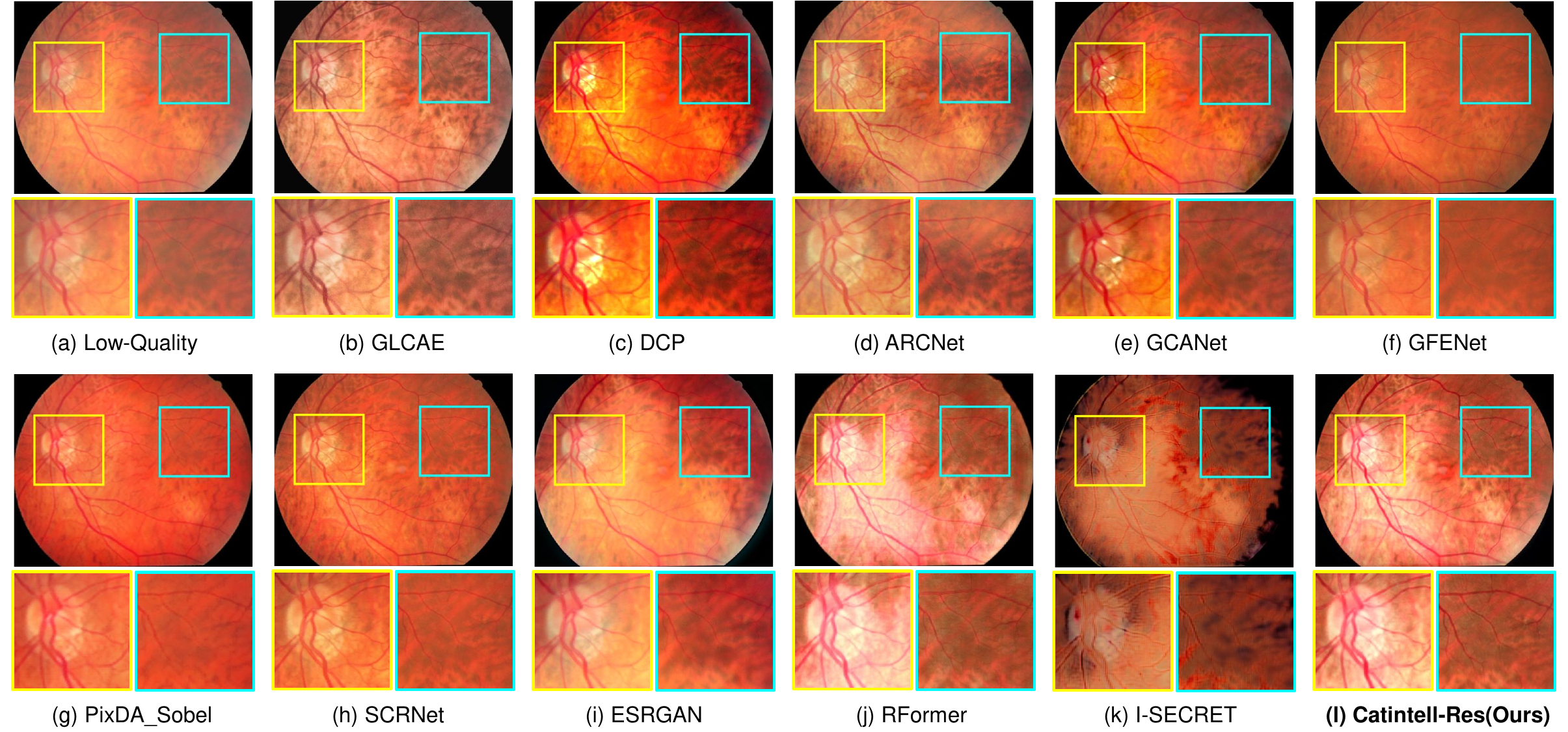}
  \caption{Restored real cataract image comparisons of Scene 2 on a test image of the Catintell Image dataset. The optic cup/disk area of the fundus image restored by Catintell-Res has clear edges of vessels. In the surrounding area, the vessels are easy to distinguish.}
  \label{fig:real_Compare2}
\end{figure*}

\begin{figure*}
  \centering
  \includegraphics[width=0.9\linewidth]{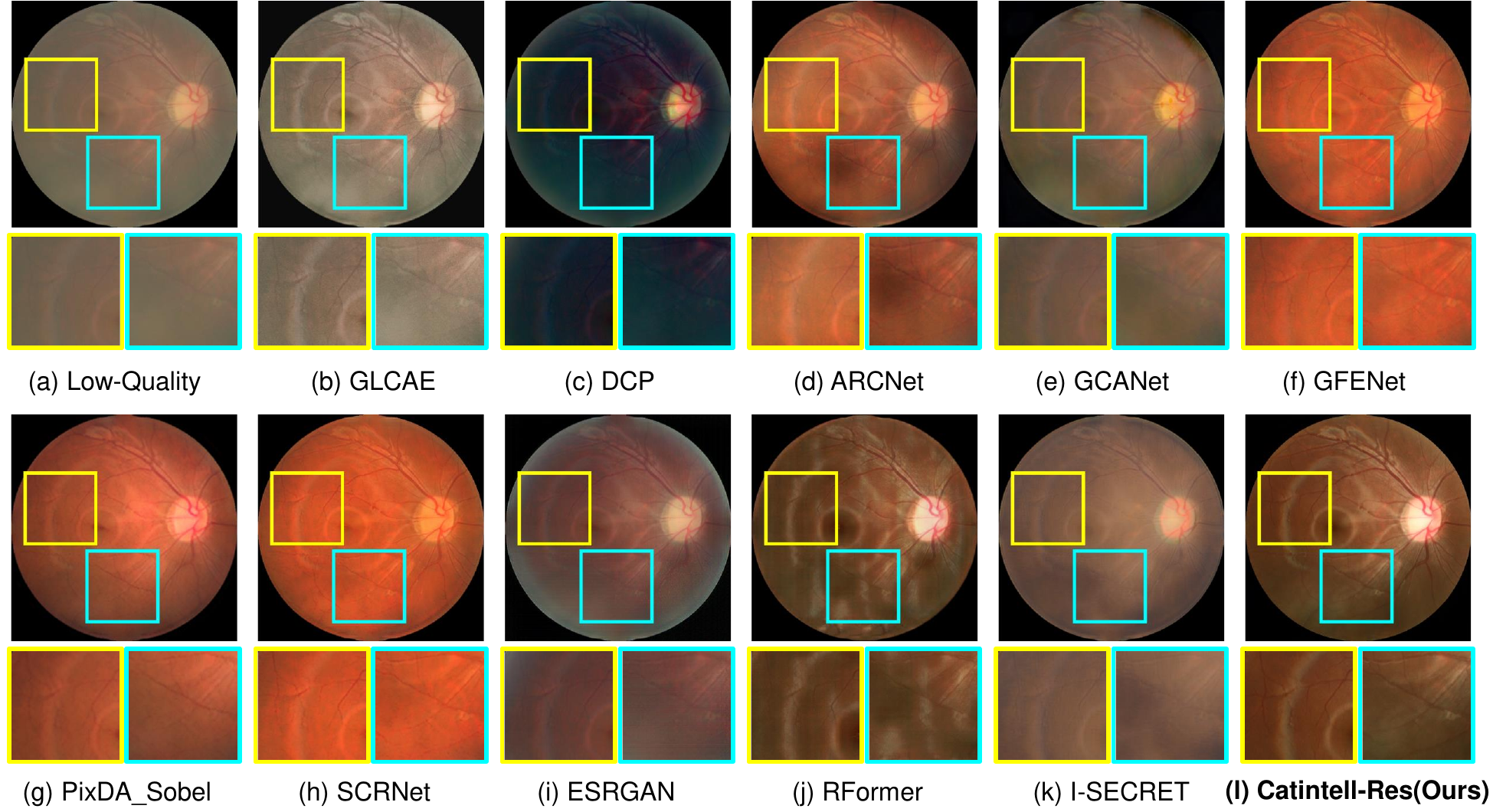}
  \caption{Restored synthesized cataract image comparisons. Catintell-Res can retain the style of the image and escalate the contrast of the whole image.}
  \label{fig:syn_Compare}
\end{figure*}

We also use the test set of the Catintell dataset to examine the restoration ability towards the real cataract image. Two samples of test results are shown in Fig\ref{fig:real_Compare1} and Fig\ref{fig:real_Compare2}. As mentioned above, the restoration branch of Catintell can work independently, this test was carried out on the real cataract images which have no corresponding clear images to compare. However, besides this visual exhibition, we also did a user study in the later section to show the results from Catintell getting the highest rating from ophthalmologists from clinical usage.

In the first real cataract image test scene, the style of the restored cataract image is retained by Catintell-Res, and the vessel details around the macular are restored and become more obvious compared to other methods and the original image. In the second scene, the optic cup/disk and surrounding area of the fundus image restored by Catintell-Res become much clearer, and the overall contrast of this image is raised.

\begin{table}
	\footnotesize
	\centering	
	\caption{Compared with SOTA methods}
	{       \begin{tabular}{c c c c}
				\toprule
				 Method & PSNR & SSIM &  Parameters(M) \\
				\midrule
				GLCAE~\cite{tian2017global} & 16.22& 0.5627 &--  \\
                Dark channel prior~\cite{8658661}  & 15.90& 0.7482 &-- \\
                \midrule
                GCANET\cite{chen2018gated} & 23.61& 0.8145 & -- \\
                ESRGAN~\cite{wang2018esrgan} & 29.47 & 0.7907  &16.72 \\
                ARCNet\cite{li2022annotation} & 19.81& 0.8709 & 54.42 \\
                
                GFENET\cite{li2023generic} & 16.77& 0.8521 & 89.30 \\
                pixDA Sobel\cite{li2021restoration} & 18.52& 0.8399 & 54.42\\
                SCRNET\cite{li2022annotation} & 15.76& 0.8568 & 89.28\\
				
				RFormer~\cite{9810184} & 22.96 & 0.6808 & 21.66 \\
                I-SECRET~\cite{cheng2021secret} & 19.19 &0.7844 & 10.85 \\
				\midrule
                \textbf{Catintell-Res(Ours)} & \textbf{39.03} & \textbf{0.9476} &  12.72\\
				\bottomrule
	\end{tabular}}
	\label{tab:CATsota}
\end{table}

\begin{table}
	\footnotesize
	\centering	
	\caption{User Study of Catintell-Res}
        \scalebox{0.8}{
			\begin{tabular}{c c c c c}
				\toprule
				 Method & \textbf{Catintell-Res(Ours)} & GLCAE &  DCP & ARCNet \\
				\midrule
				Rank Score & \textbf{99.17}& 64.67  &44.17 &78.17 \\
                \midrule
				 Method & GCANET & GFENET &  I-SECRET& Original Image \\
				\midrule
                Rank Score & 74.33 & 71.33  &34.83  &53.33\\
				\bottomrule
	\end{tabular}}
	\label{tab:Dus}
\end{table}

\begin{table}
	\footnotesize
	\centering	
	\caption{Ablation Study of Different Encoder/Decoder}
	{
			\begin{tabular}{c c c c}
				\toprule
				 Method & PSNR & SSIM &  Parameters(M) \\
				\midrule

				RRDB~\cite{wang2018esrgan} & 29.61 & 0.7649  &38.56 \\
				Swin Transformer~\cite{liu2021swin} & 24.83 &0.4150 & 15.56 \\
                ConvNeXt~\cite{liu2022convnet} & 34.97 & 0.8696 &  21.81 \\
                \textbf{Catintell-Res(Ours)} & \textbf{39.03} & \textbf{0.9476} &  12.72\\
				\bottomrule
	\end{tabular}}
	\label{tab:CATbb}
\end{table}

\subsection{Catintell-Res User Study}

After validating the restoration ability of Catintell-Res, we carried out another user study to figure out what opinions ophthalmologists hold. In the study, we provide them with eight images, which are original cataract images and images restored by GLCAE~\cite{tian2017global}, Dark channel prior~\cite{8658661}, ARCNet\cite{li2022annotation}, GCANET\cite{chen2018gated}, GFENET\cite{li2023generic}, I-SECRET~\cite{cheng2021secret}, and our Catintell-Res model. We did not label these images with methods or indicate their source. There are ten sets of these image groups, and the images are given 10,9,8,7...4,3 scores corresponding to their ranks, respectively. (This score setting means to get scores with a maximum of 100.) The average results of three experienced ophthalmologists and three young ophthalmologists are summarized in Table \ref{tab:Dus}.

The images restored by Catintell-Res are the best according to the score among these methods. Therefore, the restoration ability of Catintell-Res has proven effective and powerful, whether in quantitive experiments or user studies.

\section{Discussion}
\subsection{Generalized Cataract Restoration Ability}

Besides using the synthesized cataract and real cataract images in the Catintell Image dataset, we also test our models on the other open-source cataract dataset. The ODIR dataset is from the ODIR2019 competition\cite{odir}, which contains several kinds of fundus images of retinal diseases. We use cataract images in the training set of this dataset to validate Catintell-Res. We also collected a dataset from Kaggle named Cataract-Dataset\cite{kaggle1}. We use the cataract division of this dataset in this experiment. 

\begin{figure}
  \centering
  \includegraphics[width=1\linewidth]{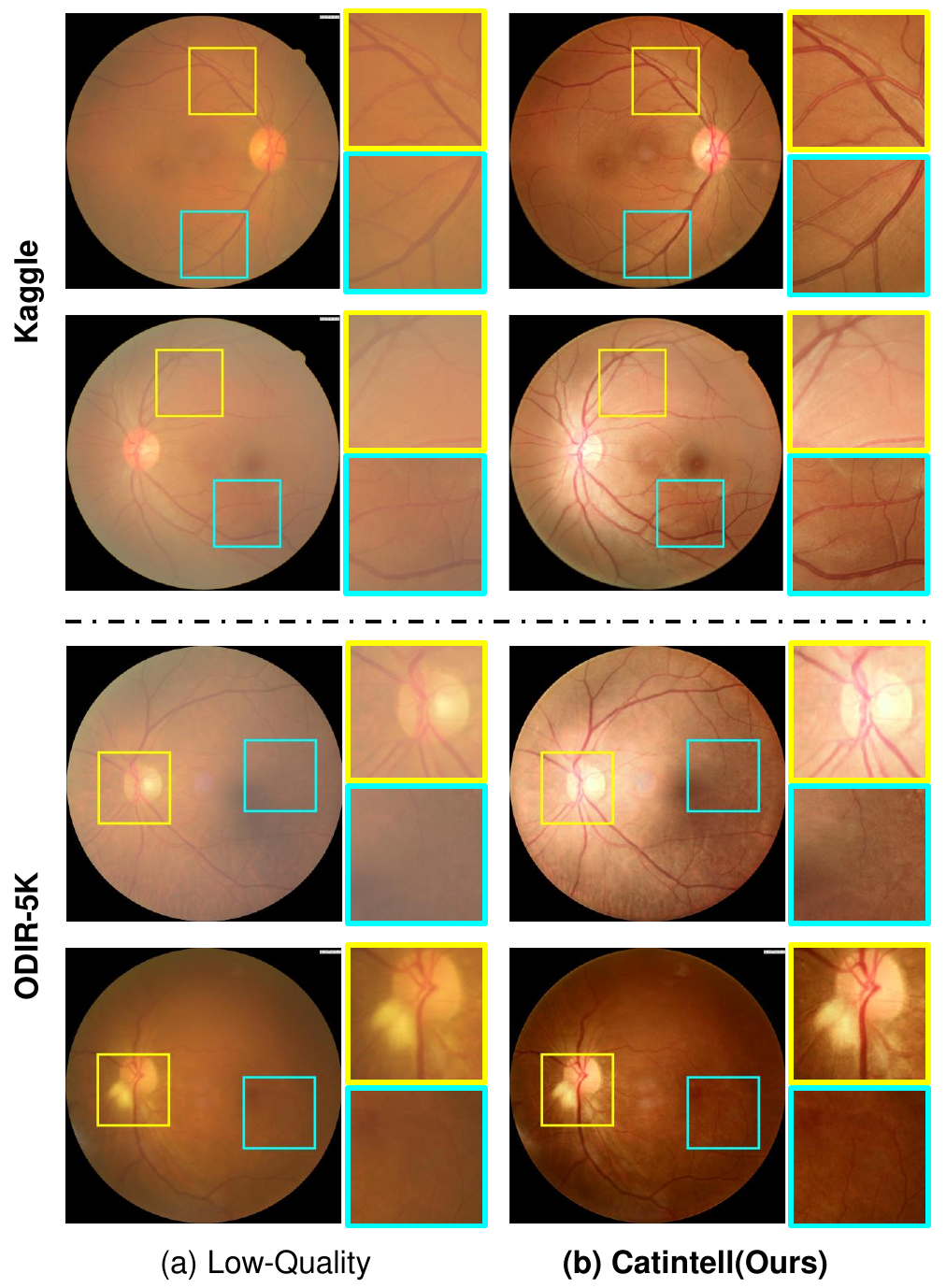}
  \caption{Restored real cataract image from external datasets. Catintell-Res has the universal ability to restore cataract images collected from other fundus cameras and sources.}
  \label{fig:External}
\end{figure}

Catintell-Res is not further retrained or modified, and the data is directly processed by the trained Catintell-Res model. The results are shown in Fig.\ref{fig:External}. We can observe from the figures that Catintell-Res has obtained universal restoration ability through the synthesized data from the Catintell Image dataset, and its ability still functions even on cataract images from other sources. In the Kaggle dataset, the macula of these fundus images is restored to be clear, and the vessels become obvious. This also suits the ODIR-5K dataset, and we can see that Catintell-Res is able to remove most of the blurry area in the real cataract images.

\subsection{Ablation Studies}

\subsubsection{Encoder/Decoder}

Though we designed a new encoder/decoder structure in Catintell-Res, we also tried other encoder structures to optimize the performance of Catintell-Res.
ConvNeXt\cite{liu2022convnet}, RRDB(residue in residue dense block) of ESRGAN~\cite{wang2018esrgan}, and W-MSA of SWIN Transformer~\cite{liu2021swin} are applied in the model of same structures to compare their performance.

The results are shown in Table \ref{tab:CATbb}. The ConvNeXt encoder/decoder has the best performance except for Catintell, which is why we optimize the encoder/decoder of Catintell with inspiration from ConvNeXt. The encoder/decoder of Catintell is optimized for image restoration and achieves the best performance among those methods.

\begin{table}
	\footnotesize
	\centering	
	\caption{Ablation Study of Different Patch Size}
	\scalebox{1}{
			\begin{tabular}{c  c  c c}
				\toprule
				 Size & PSNR & SSIM &  Parameters(M) \\
				\midrule
				128 &  35.71 & 0.9357  &12.72 \\
				192 &  38.07 &0.9279 & 12.72 \\
                
				
                \textbf{256(Catintell-Res)} & \textbf{39.03} & \textbf{0.9476} &  12.72\\
                384 & 37.51 &0.9369 & 12.72 \\
				\bottomrule
	\end{tabular}}
	\label{tab:CATpatch}
\end{table}

\subsubsection{Patch Size}

In the training process of Catintell, we use patches of size 256×256 pixels to avoid heavy computational burden. Since the patch size significantly impacts the model performance, we test different patch sizes in this section. 

When the patch size is smaller, the model can use a larger batch size during training to avoid sampling error. However, smaller patches make it difficult for the model to learn the spatial context information of the entire image and prone to overfitting, which in turn leads to a decrease in performance in the validation stage. On the other hand, when the patch size is larger, it consumes more space and forces the batch size to be reduced, and the sampling error increases, making the model hard to converge. 

As shown in Table \ref{tab:CATpatch}, the training results of the model with a patch size of 256 are the best.

\subsubsection{Depth and Width}

\begin{table}
	\footnotesize
	\centering	
	\caption{Ablation Study of Different Width}
	\scalebox{1}{
			\begin{tabular}{c c c c}
				\toprule
				  Width & PSNR & SSIM &  Parameters(M) \\
				\midrule
				16 & 36.07 & 0.9265  &3.78 \\
                \textbf{32(Catintell-Res)} & \textbf{39.03} & \textbf{0.9476} &  12.72\\
                48 & 37.11 & 0.9365 &  26.82\\
				\bottomrule
	\end{tabular}}
	\label{tab:CATwidth}
\end{table}

Since Catintell-Res uses a U-shaped structure, each encoding/decoding stage is aligned with a downsampling/upsampling, so the number of encoding/decoding stages significantly affects the network depth. 

The width of the network is determined by the projection channels of the input projection layer. With the linear increase of the projection channels, the parameters of the model increase quadratically.

To obtain the optimal number of encoding/decoding stages and network width, we conduct the following experiments on the Catintell-Res model, keeping the rest of the structure unchanged and only changing the number of encoding/decoding stages or width to verify its impact on performance. The results are summarized in Table \ref{tab:CATstage} and \ref{tab:CATwidth}, and the model with four stages and a width of 32 has the best performance. Therefore, this width and depth combination is used in the Catintell-Res model to obtain the best model performance.

\begin{table}
	\footnotesize
	\centering	
	\caption{Ablation Study of Different Depth}
	\scalebox{1}{
			\begin{tabular}{c c c c}
				\toprule
				 Depth & PSNR & SSIM &  Parameters(M) \\
				\midrule
				3 & 36.78 & 0.9346  &3.711 \\
                \textbf{4(Catintell-Res)} & \textbf{39.03} & \textbf{0.9476} &  12.72\\
                5 & 34.07 & 0.9005  &46.27 \\
				\bottomrule
	\end{tabular}}
	\label{tab:CATstage}
\end{table}

\subsection{Limitation}
Though Catintell-Res has obtained universal restoration ability, it can not process images with severe blur. When fundus images are collected, there is some reason that their quality is not guaranteed. To be more specific, some images suffer from wrong illumination, whether too high or too low. And some may be blocked by eyelids or iris. All of those abnormal images can be named degradation images. For those images with severe degradation, there is no sign of vessels to assist Catintell-Res in escalating image quality. Therefore, Catintell-Res can not handle these images or generate whole images through a little undegraded area. 


Recently, diffusion models have attracted some interest from researchers in the image-to-image translation field. We also regard diffusion models like works from Rombach $\textit{et al}.$\cite{rombach2021highresolution} and Su $\textit{et al}.$\cite{su2022dual} as good solutions for both cataract image synthesis and restoration. However, the diffusion models could occupy a massive amount of GPU RAM while training, which is sometimes over 40GB in practice and much higher than the inference process. This RAM burden is too heavy for our GPU to train a diffusion model. 

Moreover, suppose the diffusion steps are high or the latent feature size is small. In that case, the generation ability of the diffusion model is too strong to retain enough fidelity for medical usage, and fake focus may be generated due to this. Therefore, we choose not to use diffusion models in our work for now, but, still, diffusion models are of great potential in medical image processing which we plan to exploit in the future.

We plan to:
\begin{itemize}
    \item [1] Enlarge the range of images collected in the Catintell Image dataset to elevate the generating ability of Catintell-Res for a more extensive range of low-quality images.
    \item [2] Modify the structure of Catintell-Syn to make it able to generate more kinds of degraded images.
    \item [3] Transfer the Catintell models to other medical image tasks to extend their application.
    \item [4] Apply lightweight diffusion models on fundus image restoration and optimize Catintell models.
\end{itemize}

\section{Conclusion}

In this paper, we address the problems in cataract image restoration through a new synthesizing and restoration method, Catintell. 
Before our method, there was much difference between conventional simulated and real cataract images; the quality of restored cataract images was not high enough. Our method, Catintell-Syn, uses fully unsupervised data to generate paired cataract-like images with realistic style and texture and successfully alleviates the lack of paired images. Based on the synthetic images, we developed Catintell-Res to restore real cataract images. The structure of these models is optimized for fundus images, and we also added the loss function expertized for ophthalmology in the training stage.
Then, we carried out user studies and quantitive experiments for Catintell models. The results show that Catintell achieves remarkable performance in both synthesizing cataract-like data and restoring real cataract data.
The generalization performance of Catintell-Res is verified by real cataract images from various external datasets. We plan to open Catintell models for research and clinic utilization and hope this model can help ophthalmologists with their work in the future.

\section{Acknowledgments}

This work is supported by the Science and Technology Innovation Committee of Shenzhen-Platform and Carrier (International Science and Technology Information Center) \& Shenzhen Bay Lab under KCXFZ20211020163813019 and by the National Natural Science Foundation of China under 82000916.

The authors declare no other conflict of interest.  

This study was performed in line with the principles of the Declaration of Helsinki. The Ethics Committee of Beijing Tongren Hospital, Capital Medical University granted approval and supervised the research process. 



\printcredits

\bibliographystyle{cas-model2-names}
\bibliography{main}

\end{document}